\documentclass[10pt,abstract]{sensys-proc}

\usepackage{textcomp}
\usepackage[compact]{titlesec}
\usepackage{graphicx}
\usepackage{url}
\usepackage[font=small,labelfont=bf, skip=0pt]{caption}
\begin{document}

\title{Poster Abstract: Bits and Watts: Improving energy disaggregation performance using power line communication modems}

\numberofauthors{1} %  in this sample file, there are a *total*
\author{Nipun Batra$^1$, Manoj Gulati$^1$, Puneet Jain$^1$,  Kamin Whitehouse$^2$, Amarjeet Singh$^1$\\
\small $^1$Indraprastha Institute of Information Technology Delhi, India ~\{nipunb,~manojg,~puneet13150,~amarjeet\}@iiitd.ac.in\\
\small $^2$ University of Virginia ~whitehouse@virginia.edu
}

\crdata{978-1-4503-3144-9}
\doidata{10.1145/2674061.2675039}
\conferenceinfo{To appear in BuildSys'14,} {November 5--6, 2014, Memphis, TN, USA.}
\CopyrightYear{2014}

% There's nothing stopping you putting the seventh, eighth, etc.
% author on the opening page (as the 'third row') but we ask,
% for aesthetic reasons that you place these 'additional authors'
% in the \additional authors block, viz.
%\additionalauthors{Additional authors: }
%\date{19 August 2014}
% Just remember to make sure that the TOTAL number of authors
% is the number that will appear on the first page PLUS the
% number that will appear in the \additionalauthors section.

\maketitle
\begin{abstract}
Non-intrusive load monitoring (NILM) or energy disaggregation, aims to disaggregate a household's electricity consumption into constituent appliances. More than three decades of work in NILM has resulted in the development of several novel algorithmic approaches. However, despite these advancements, two core challenges still exist: i) disaggregating low power consumption appliances and ii) distinguishing between multiple instances of similar appliances. These challenges are becoming increasingly important due to an increasing number of appliances and increased usage of electronics in homes. Previous approaches have attempted to solve these problems using expensive hardware involving high sampling rates better suited to laboratory settings, or using additional number of sensors, limiting the ease of deployment. In this work, we explore using commercial-off-the-shelf (COTS) power line communication (PLC) modems as an inexpensive and easy to deploy alternative solution to these problems. We use the reduction in bandwidth between two PLC modems, caused due to the change in PLC modulation scheme when different appliances are operated as a signature for an appliance. Since the noise generated in the powerline is dependent both on type and location of an appliance, we believe that our technique based on PLC modems can be a promising addition for solving NILM. 
%We also release our initial data set for public use.
\end{abstract}

%\category{H.4}{Information Systems Applications} Miscellaneous

%\keywords{energy disaggregation; non-intrusive load monitoring; smart meters}

\section{Introduction}
\label{intro}
More than three decades ago, George Hart and his team~\cite{hart_1992} laid the foundation of non-intrusive load monitoring. While the problem continues to intrigue researchers, it has undergone several important changes in these three decades. From a hardware perspective, smart meters are increasingly more affordable and easy to install. Traditional electrical appliances in homes are increasingly replaced by their more efficient electronic counterparts. As a result, homes generally contain far more appliances than they used to three decades ago. Many of these electronic appliances such as phone chargers consume low power. Additionally, multiple instances of the same appliance are increasingly common these days.  

\begin{figure}
\vspace{-6pt}
\centering \includegraphics[scale=0.8]{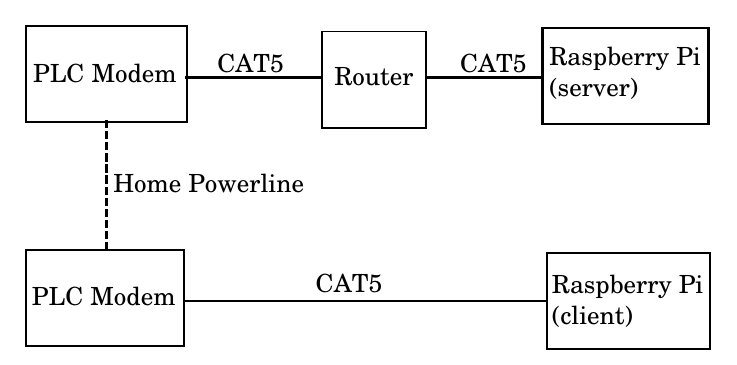}
\vspace{-3pt}
\caption{\small{Experimental setup: We measure the change in bandwidth between two Raspberry Pis when appliances are turned on/off.}}
\label{setup}
\end{figure}

Previous NILM approaches have mostly focused on high power consuming appliances and thus express their limitations in disaggregating i) low power consuming appliances and ii) similar appliances. However, a few approaches for solving these problems have been proposed in the past. Gupta et al.~\cite{electrisense} and Gulati et al.~\cite{emisense} use a single point based high speed sampling hardware to capture the EMI noise from switched mode power supply based appliances and disaggregate them. While the solution promises disaggregating low power appliances and multiple instances of similar appliances, it is limited by expensive hardware, data management issues arising due to high frequency data collection and improved EMI filters nulling out EMI noise. 
%Completely tangential to this work, Barker et al.~\cite{barker} recently proposed smart outlets based energy disaggregation. However, most outlets used currently are dumb and replacing with newer smart outlets requires significant deployment effort. 
%Rowe et al.~\cite{contactless} devise an EMF sensor which is located close to an appliance and can detect power changes from a few inches away. 

In contrast, we aim to explore COTS PLC modems to aid energy disaggregation. In essence, our proposed approach involves placing two PLC modems at different outlets in a home and observing the bandwidth change between them in correlation to power change in the home. We now discuss our choice of PLC modems.

\section{Why PLC?}
\label{intuition}
%``To know and not to do is really not to know." 
Our quest to look into PLC modems as a possible solution to the problems discussed in Section \ref{intro} began an year earlier, based on a work published in Buildsys 2013, which presented the performance of PLC modems for green building applications~\cite{roy2013performance}. This work demonstrated the degrading bandwidth performance of modern day PLC modems when appliances (motor based in particular) are used in the vicinity. Murty et al.~\cite{murty} suggest that this degrading performance is due to the change in PLC modulation scheme when different appliances are operated. They further suggested that reactive appliances are more prone to cause performance degradation in comparison to resistive appliances. Furthermore, they show the variation of bandwidth with length of cable between the modems. While these works had studied the impact of appliance activity and wire length on PLC with an aim of characterising PLC performance, we utilise the same techniques to do the opposite, characterise appliance usage based on PLC performance. With this in mind, we feel that two instances of the same appliance, which are located at physically different parts of a home will cause different PLC performance degradation. Further, we chose PLC modems due to their inexpensive and plug-and-play nature.
%We now discuss our experimental setup.

\section{Experimental setup}

We tried to emulate the experimental setup as per the papers discussed in Section \ref{intuition}. We placed two TPLink HomePlug AV PLC modems\footnote{\url{http://www.tp-link.com/lk/products/details/?model=TL-PA2010}} at separate outlets in a home and did a continuous data transmission between two Raspberry Pis using the \texttt{iperf} Unix utility. We recorded the instantaneous bandwidth, data transmission (in bytes) and timestamp every second on the client Raspberry Pi. Figure \ref{setup} shows our experimental setup. It must be noted that no other computing device is on this network and thus the network performance is not subject to external disturbance.

\section{Evaluation}

\begin{figure}
\vspace{-6pt}
\centering \includegraphics[scale=0.9]{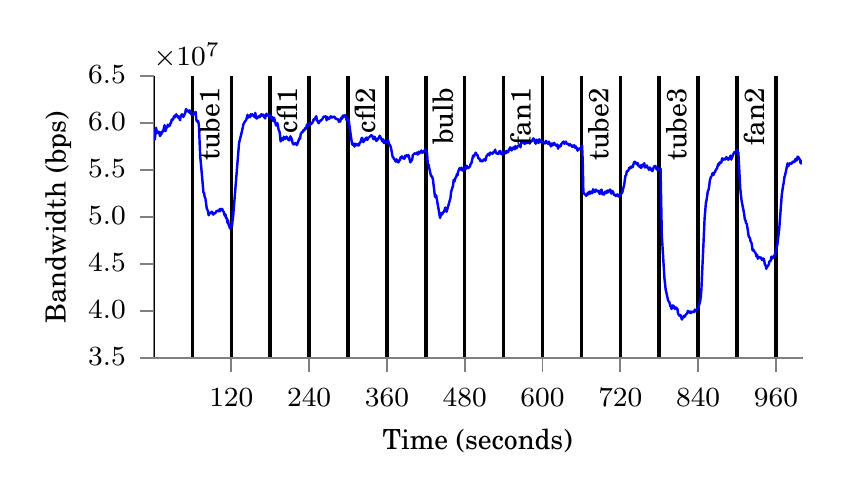}
\vspace{-6pt}

\caption{\small{Reduction in bandwidth observed due to the operation of different appliances. tube2, tube3 and fan2 are located in the same room and the remaining in a different room.}}
\label{result1}

\vspace{-6pt}
\end{figure}

We ran our experiment in a single storey home in Delhi, India. We turned off all appliances (no-load settings) in the home before starting our controlled experiment. Our evaluation involved turning on an appliance, leaving it on for 60 seconds before turning it off and turning on the next appliance after a gap of 60 s. Figure \ref{result1} shows the observed bandwidth when different household appliances were used. We can observe that tube1, tube2 and tube3 which are located in different locations in the home cause distinct reduction in bandwidth. However, cfl1 and cfl2 which are located very close to each other cause similar bandwidth reduction. We thus believe that our technique can distinguish between multiple instances of the same appliance if they are sufficiently physically separated.

\begin{figure}
\vspace{-6pt}
\centering \includegraphics[scale=0.9]{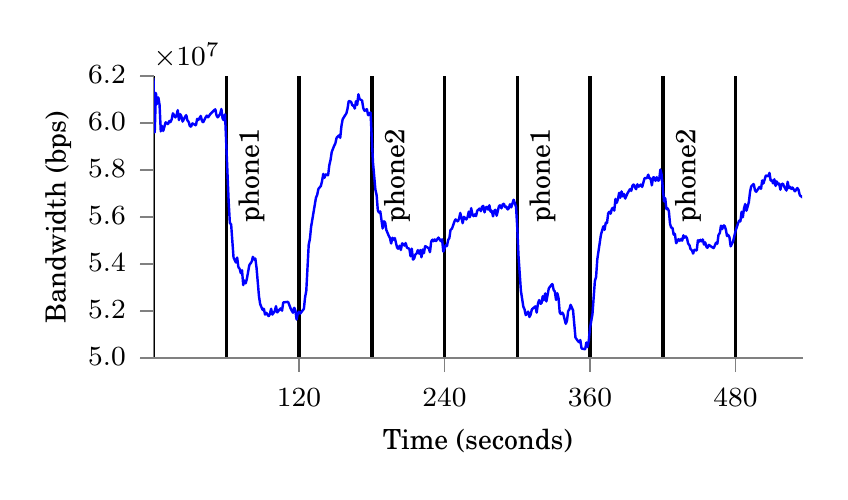}
\caption{\small{Drop in bandwidth when two phone chargers were operated.}}
\label{result2}
\vspace{-6pt}
\end{figure}

Figure \ref{result2} shows the drop in bandwidth observed when two phone chargers, which consume 5 Watts or less, are used at different locations. Based on this observation, we believe that our technique may be able to disaggregate low power appliances. Further, we repeated our experiments to validate that the drop in bandwidth for an appliance remains reasonably constant over time. However, we must also point out that we observed fluctuation in no-load characteristics over time (as shown in Figure \ref{result1} when the no-load bandwidth drop around 400 seconds).

\section{Conclusions and future work}
In this work we presented a PLC modem based technique to address two NILM problems: distinguishing multiple instances of similar appliances and disaggregating low power consuming appliances. Based on our initial analysis, we believe that our system can addresses these problems. In the future, we would like to analyse how the simultaneous operation of appliances affects PLC bandwidth. Further, we would also like to repeat this experiment across more homes to validate our findings. Our current experiments were done in a home in New Delhi, where the electrical grid is considered unreliable~\cite{iawe}. In the future, we wish to do these experiments under different grid stability settings. Eventually, we would like to use PLC based disaggregation together with conventional NILM using an NILM toolkit such as NILMTK~\cite{nilmtk}. 

\bibliographystyle{abbrv}
\begin{small}
\bibliography{sigproc}  % sigproc.bib is the name of the Bibliography in this case
\end{small}
% ACM needs 'a single self-contained file'!
%\balancecolumns

% That's all folks!

\end{document}